\documentclass[prl,twocolumn,showpacs,superscriptaddress,floatfix]{revtex4}
\usepackage{graphicx}
\usepackage{dcolumn,float}
\usepackage{amsmath,amssymb}
\usepackage{subfigure}

\begin{document}

\title{Quantifying entanglement in multipartite conditional states of open quantum systems by measurements of their photonic environment}

\author{Juan Diego Urbina}
\affiliation{Institut f\"ur Theoretische Physik, Universit\"at Regensburg, D-93040 Regensburg, Germany}
\author{Walter T. Strunz}
\affiliation{Institut f\"ur Theoretische Physik, Technische Universit\"at Dresden, D-01062 Dresden, Germany}
\author{Carlos Viviescas}
\affiliation{Departamento de F\'{\i}sica, Universidad Nacional de Colombia, Carrera 30 No.45-03, Bogota D.C., Colombia}

\begin{abstract}

A key lesson of the decoherence program is that information flowing out from an open system is stored in the quantum state of the surroundings. Simultaneously, quantum measurement theory shows that the evolution of any open system when its environment is measured is nonlinear and leads to pure states conditioned on the measurement record. Here we report the discovery of a fundamental relation between measurement and entanglement which is characteristic of this scenario. It takes the form of a scaling law between the amount of entanglement in the conditional state of the system and the probabilities of the experimental outcomes obtained from measuring the state of the environment. Using the scaling, we construct the distribution of entanglement over the ensemble of experimental outcomes for standard models with one open channel and provide rigorous results on finite-time disentanglement in systems coupled to non-Markovian baths. The scaling allows the direct experimental detection and quantification of entanglement in conditional states of a large class of open systems by quantum tomography of the bath.
\end{abstract}


\maketitle

Arguable, a defining feature of a \emph{real} quantum system is the nature of its unavoidable interactions with its surroundings. It is well known that they are determinant of the manners the system can evolve, being even responsible, in occasions, of the total loss of the system quantum character \cite{zurek}. Less discussed, however, is the fact that they also dictate the ways a system can be observed and information can be extracted from it. Thus, for instance, a system can be indirectly probed by performing a projective measurement on its environment. Ignore the outcome of this measurement, and the information lost implies that a pure state of the system prior observation is transformed into a statistical mixture \cite{meas1,meas2}. If on the other hand, the result of the measurement is physically stored with full fidelity the state of the central system immediately after measurement is again pure, yet conditioned on the particular, stochastic outcome: the state, representing our knowledge about the system, ``collapses" \cite{open1,open2}. Conditional states not only have an intrinsic importance as the outcome of physical measurements. They provide also an exact representation of the unconditional dynamics, with implications for the field of quantum information \cite{nuev0,nuev1,nuev2} where characterizing the amount and kind of non-classical correlations in unconditional states of open systems remains a main theoretical challenge \cite{qi,qi1}. In this order of ideas, a proper understanding of the interplay between measurements on the bath and entanglement properties of conditional states, to our knowledge a problem not yet explored, is clearly necessary.

Under the only assumption of a strictly local dynamics, in the present article we derive a scaling law relating the probability distribution of experimental outcomes from measurements on the environment, as given by the Born rule, to the amount of multipartite entanglement of the pure conditional state of an open quantum system. As we demonstrate, our findings are universal: They are independent of the number and dimensions of the constituent subsystems and hold for any of the so-called G-invariant measures of entanglement \cite{scal4,scal2}, e.g., Wootter's concurrence for two-qubit systems \cite{concu1}, characterized by their invariance under local invertible transformations. Notably, these results allow for a characterization of multipartite entanglement in the presence of both Markovian and non-Markovian environments, thus making contact with state-of-the-art experimental techniques \cite{exp}. Furthermore, we show that our scaling law suggests a protocol for a direct measurement of entanglement in a large class of systems, including all Markovian and all non-demolition baths: As it turn out, the inverse numerical value of the Husimi function of the bath is a measure of the amount of entanglement in the conditional state. 

Our work is partially motivated by recent results in the study of non-classical correlations in unconditional states, where a conservative approach  based on explicit calculation of the entanglement measures in the unconditional mixed state faces tremendous difficulties due to an intrinsic high dimensional optimization problem \cite{cond1,cond2}. The severity of this hurdle has motivated intense research in the general (algebraic) aspects of the interplay between unconditional time evolution and entanglement measures. An important step in this direction was the discovery of a {\it scaling} property of  the G-invariant measures \cite{scal1,scal2}, and its experimental verification for two-qubit systems with one open channel \cite{scal3}. 

However, the scaling property of unconditional states critically depends on the linearity of the evolution \cite{scal2} through its Kraus representation \cite{qi}. Therefore, an extension of such algebraic properties to the conditional case with its intrinsic nonlinear dependence on the initial state requires both different  mathematical methods and physical concepts. Our results fill this gap at a moment when there is an increasing interest on schemes with a less restrictive role for the environment, where the flow of information from the system to the bath takes an active part \cite{nuev0,nuev1,nuev2,nuev2.5,nuev3,nuev4,nuev5,nuev6}.

\subsection{Open systems in photonic environments}

We consider a central-system-plus-bath approach \cite{open1,open2} where a multipartite central system (CS) made of $N$ subsystems with dimensions $d^{(1)},\ldots,d^{(N)}$ is coupled to a collection of independent photonic baths (channels) acting locally on each subsystem. If initially pure, after transformation to the interaction picture, the state of the combined system is given at time $t$ by 
\begin{equation}
\label{eq:Psi}
|\Psi(t)\rangle={\cal T}{\rm e}^{-\frac{i}{\hbar}\int_{0}^{t} (\hat{H}_{CS}(s)+\hat{H}_{I}(s))ds}|\phi(0)\rangle \otimes |0\rangle.
\end{equation}
In the above equation $|\phi(0)\rangle$ is the initial state of the CS and $|0 \rangle$ the ground state of the bath, ${\cal T}$ stands for the time ordering operator, $\hat{H}_{CS}(s)$ is the CS Hamiltonian acting locally on the subsystems, and
\begin{equation*}
\hat{H}_{I}(s)=i\sum_{k,\lambda}g_{k,\lambda}\left(\hat{J}_{k}\hat{a}_{k,\lambda}^{\dagger}{\rm e}^{i \omega_{k,\lambda}s}-\hat{a}_{k,\lambda}\hat{J}_{k}^{\dagger}{\rm e}^{-i \omega_{k,\lambda}s}\right),
\end{equation*}  
accounts for the interaction between CS and bath through local traceless operators $\hat{J}_{k}$, with $g_{k,\lambda}$ the coupling strength of the system and the $\lambda$-th mode to the $k$-th channel, where photons of frequency $\omega_{k,\lambda}$ are created or annihilated by $\hat{a}_{k,\lambda}^{\dagger}$ and $\hat{a}_{k,\lambda}$, respectively. Memory effects in this approach are characterized by the correlation functions
\begin{equation*}
\label{eq:resa}
\alpha_{k}(s)=\sum_{\lambda}|g_{k,\lambda}|^{2}{\rm e}^{-i \omega_{k,\lambda}s}=\int_{0}^{\infty}I_{k}(\omega){\rm e}^{-i \omega s}d\omega, 
\end{equation*}
where $I_{k}(\omega)$ is the spectral density of the bath's $k$-th channel (see Supplementary Information).

\subsection{Universal scaling of multipartite entanglement} 

Following \cite{qt0,qt1,qt2}, the total state $|\Psi(t)\rangle$ in equation (\ref{eq:Psi}) can be always written as
\begin{equation}
\label{eq:Psizeta}
|\Psi(t)\rangle={\cal Z}^{-1}\int{\rm e}^{-\frac{|a|^{2}}{2}}|\psi(a,t)\rangle \otimes |a\rangle da,
\end{equation}
where $|\psi(a,t)\rangle$ is the (unnormalized) state of the system \emph{relative} to the bath (normalized) coherent state $|a\rangle$, ${\cal Z}$ is a normalization constant, which will drop out from all the final results, and we define $da:=\prod_{k,\lambda}d\Re{a}_{k,\lambda}d\Im{a}_{k,\lambda}$ with $|a|^{2}=\sum_{k,\lambda}|a_{k,\lambda}|^{2}$.

If at time $t>0$ a generalized measurement of the bath is performed, the probability $P(a,t)$ that we obtain $|a\rangle$ is \cite{qt1,meas2}
\begin{equation}
\label{eq:Ps}
\frac{P(a,t)}{P(a,0)}=\langle \psi(a,t)|\psi(a,t)\rangle\,,
\end{equation} 
with $P(a,0)={\cal Z}^{-1}{\rm e}^{-|a|^{2}}$, and the state of the central system collapses to the conditional state 
\begin{equation}
\label{eq:phScale}
|\phi(a,t)\rangle=\langle \psi(a,t)|\psi(a,t)\rangle^{-1/2}|\psi(a,t)\rangle\,.
\end{equation} 

To proceed further, we notice that the unnormalized conditional state $|\psi(a,t)\rangle$ can be related to the physical initial state of the CS by means of \cite{qt0}
\begin{equation}
\label{eq:psUp}
|\psi(a,t)\rangle =\hat{\cal U}(a,t)|\phi(0)\rangle
\end{equation}
with
\begin{equation*}
\hat{\cal U}(a,t) ={\cal T}{\rm e}^{\int_{0}^{t}\left[-\frac{i}{\hbar}\hat{H}_{\rm CS}+\sum_{k}\left(z^{*}_{k}(s)\hat{J}_{k}-\hat{J}^{\dagger}_{k}\hat{{\cal O}}_{k}(s)\right)\right]ds} \,,
\end{equation*}
a non-unitary propagator depending on the measurement outcomes $\{a\}$ through $z_{k}(s)=\sum_{\lambda}g_{k,\lambda}a_{k,\lambda}{\rm e}^{-i \omega_{k,\lambda}s}$ and $\hat{{\cal O}}_{k}(s)=\int_{0}^{s}\alpha_{k}(s-s')\hat{O}_{k}(a_{k},s,s')ds'$. Although in general the operators $\hat{O}_{k}(a_{k},s,s')$ are defined only perturbatively for each particular choice of $I_{k}(\omega)$ and $\hat{J}_{k}$, exact closed expressions are available for several cases of physical interest \cite{qt0,qt1,qt2}.  

Having a linear relation between $|\psi(a,t)\rangle$ and $|\phi(0)\rangle$, we can turn to the main issue of this work: the amount of entanglement $G(|\phi(a,t)\rangle)$ in the conditional state $|\phi(a,t)\rangle$ as quantified by a $G$-invariant measure. As we now show, the answer to this question follows smoothly from two defining properties of $G$ \cite{scal2}: (i) It is invariant under local linear transformations with unit determinant. (ii) It is homogeneous of degree two, i.e., $G(u|\phi\rangle)=|u|^{2}G(|\phi\rangle)$ for all complex $u$. Indeed, from property (ii) and Eqs.~\eqref{eq:phScale},\eqref{eq:psUp}, the entanglement of $|\phi(a,t)\rangle$ can be written as
\begin{equation*}
\label{eq:carlos}
G(|\phi(a,t)\rangle)=\langle \psi(a,t)|\psi(a,t)\rangle^{-1}G(\hat{\cal U}(a,t)|\phi(0)\rangle).
\end{equation*}
We now use property (i), in conjunction with equation~\eqref{eq:Ps}, to obtain the central result of this paper,
\begin{equation}
\label{eq:resI}
x(a,t) := \frac{G(|\phi(a,t)\rangle)}{G(|\phi(0)\rangle)} = f(a,t)\frac{P(a,0)}{P(a,t)},
\end{equation}
where the scaling function
\begin{equation*}
\label{eq:resf}
f(a,t)={\rm e}^{-\sum_{k}\frac{2\Re}{d^{(k)}}\int_{0}^{t}\int_{0}^{s}\alpha_{k}(s-s'){\rm Tr}\left[\hat{J}_{k}^{\dagger}\hat{O}_{k}(a_{k},s,s')\right]ds ds'}
\end{equation*}
is independent of the functional form of $G(|\phi\rangle)$; a consequence of the use of the identity $\text{det}{\cal T} {\rm e}^{\int_{0}^{t}\hat{A}(s)ds}={\rm e}^{\int_{0}^{t}{\rm Tr }\hat{A}(s)ds}$ and the fact that the Hamiltonians and baths act locally in the subsystems (see Methods section). Equation \eqref{eq:resI} provides us with an explicit scaling between the amount of entanglement in the conditional state of the central system and the probability distribution of the outcomes of the measurement on the bath. It expresses universality of the multipartite entanglement in the sense that {\it for a given outcome of the measurement, all normalized G-invariant measures of multipartite entanglement in the conditional state have exactly the same numerical value}. Within the standard system-plus-bath approach for local systems, our equation~(\ref{eq:resI}) is completely general. 

To illustrate the relevance of scaling law \eqref{eq:resI}, in the remainder of the paper we use it to address some crucial questions in the actual theory of multipartite entanglement.

\subsection{Distribution of multipartite entanglement}

Perhaps the most pertinent question at this point is: what is the probability of having a given amount of entanglement in a system just after its photonic environment has been measured? Let us remark that the common approach of using uniform distributions for $|\phi\rangle$ \cite{rmt1,rmt2} does not hold here. Once the measurement process is specified, i.e., a choice for the measurement outcomes is made, unitary invariance as a guiding principle is replaced by the specific quantum-mechanical distribution of outcomes $P(a,t)$, as given by the Born rule. Accordingly, the distribution $P_{G}(x,t)$ of the normalized entanglement $x(a,t)$ in the physical ensemble of pure and random conditional states of the CS must be calculated from
\begin{equation}
\label{eq:Pofx}
P_{G}(x,t)= \int\! P(a,t)\delta\left(x-x(a,t)\right) da\,,
\end{equation}
which can be evaluated using the scaling rule \eqref{eq:resI} and is the same for all $G$-invariant measures. In particular, its first moment, i.e., the mean entanglement $\bar{x}(t)$, is given by
\begin{equation}
\label{eq:barx}
\bar{x}(t)=\int_{0}^{\infty}\! P_{G}(x,t)\,x\, dx = \int\! P(a,0)f(a,t)\,da.
\end{equation}

To exemplify our argument we consider a specific and experimentally relevant case \cite{scal3}: a multipartite CS where only one qubit is coupled to a single open channel ($k=$``ch''). Hence, the information about the initial state $|\phi(0)\rangle$ is fully encoded in the initial reduced density matrix of the coupled qubit,
\begin{equation*}
\hat{\rho}_{\rm ch}(0)=\left(
    \begin{array}{cc}
            \rho_{11} &  \rho_{10}  \\
            \rho_{10}^{*} &  \rho_{00}
          \end{array}
        \right).
\end{equation*}

\begin{figure}[t]
\begin{center}
\includegraphics[scale=0.625]
{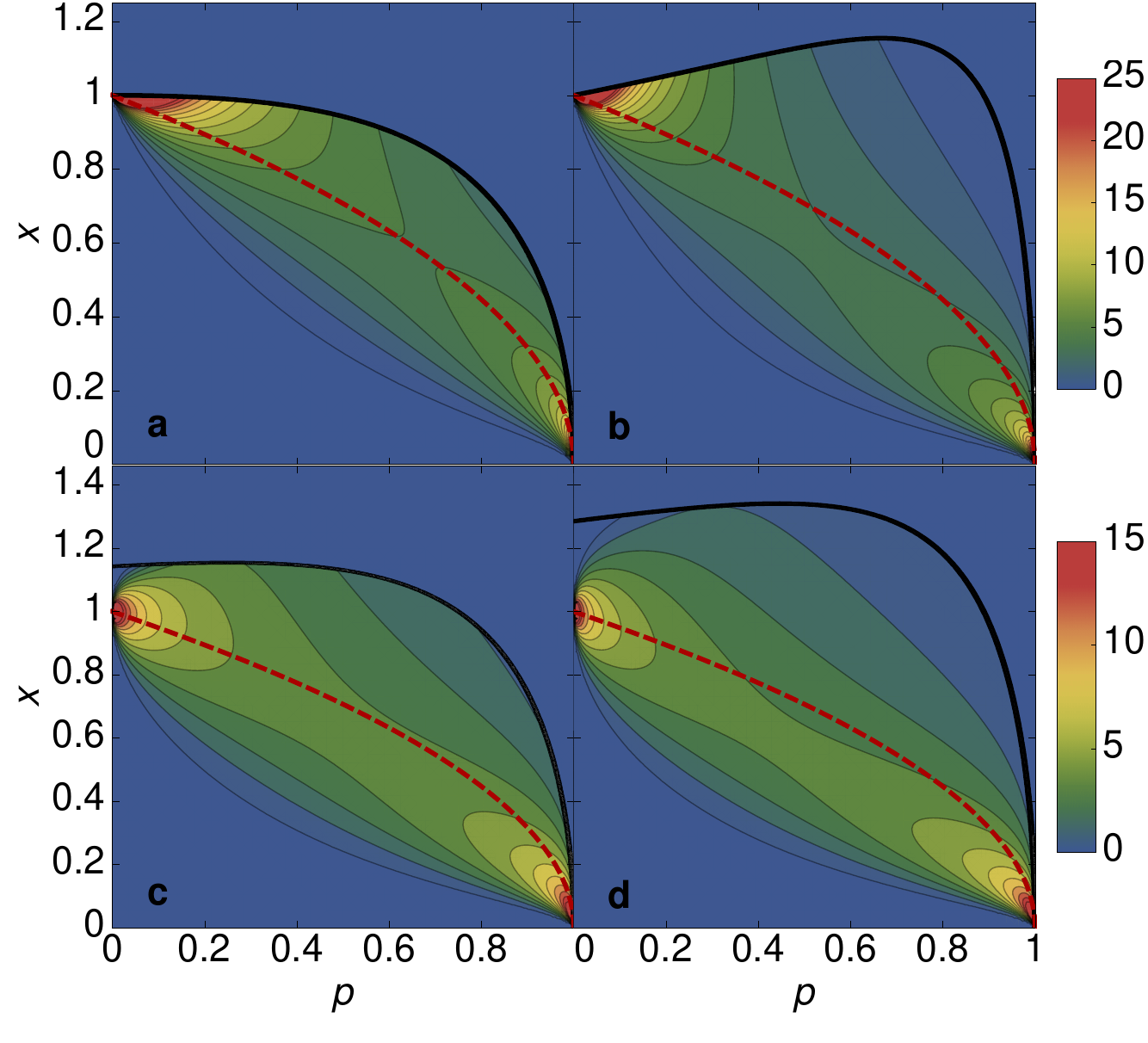}  
\caption{\textbf{Probability distribution of entanglement.} Probability distribution of scaled multipartite entanglement in the CS over the ensemble of experimental outcomes $\{a\}$ of an amplitude damping bath for different initial states: \textbf{a,} $\rho_{11}=1/2$, $\rho_{10}=0$, \textbf{b,} $\rho_{11}=3/4$, $\rho_{10}=0$, \textbf{c,} $\rho_{11}=1/2$, $\rho_{10}=1/4$, \textbf{d,} $\rho_{11}=1/2$, $\rho_{10}=1/3$. The figure is the same for any G-invariant measure of a system of qudits if only a single qubit without internal dynamics is coupled to a zero temperature Markovian bath. The black continuous line shows the maximum possible value of entanglement $x^{\text{max}}(p)$ while the red dashed line corresponds to the mean entanglement $\bar{x}(p)$.} 
\label{fig:fig1}
\end{center}
\end{figure}

Let us start by considering a Markovian bath at zero temperature where $\hat{J}=\sqrt{\gamma}\hat{\sigma}_{-}=\hat{O}(a,s,s')$ \cite{qt1}, with $\hat{\sigma}_{-}$ the two-level deexcitation operator, $\gamma$ the  decay rate, and  $ I(\omega)=\pi^{-1}$.  In order to focus on decoherence effects, we assume frozen internal dynamics. In Fig.~\ref{fig:fig1} we show $P_{G}(x,p)$ in terms of $ \hat{\rho}_{\rm ch}(0)$ and the scaled time $p(t)=1-{\rm e}^{-\gamma t}$ (see Supplementary Information). In all cases we observe that the entanglement distribution becomes broader as time increases to finally collapse asymptotically when entanglement from CS disappears. The sharp boundary exhibit by $P_{G}(x,p)$ (black continuous line) is given by the condition 
\begin{equation*}
x^{\rm max}(p)=\frac{\sqrt{1-p}}{\rho_{11}}\left(\frac{1-\rho_{11}p}{\rho_{11}}-\frac{|\rho_{10}|^{2}}{\rho_{11}^{2}}\right)^{-1}.
\end{equation*}
Thus, for maximally entangled states, i.e., states with ${\rm Tr} \hat{\rho}_{\rm ch}^{2}(0)=1/2$, the maximum possible value of entanglement $x^{\rm max}(p)$ is strictly decreasing with $x^{\rm max}(p)<1$, as can be seen in panel 1a. However, for initial states with ${\rm Tr} \hat{\rho}_{\rm ch}^{2}(0)>1/2$, there are regions in which $x^{\rm max}(p)$ can increase even up to $x^{\rm max}(p)>1$, as displayed in panel 1b. This behaviour should be contrasted with the evolution of the averaged normalized entanglement $\bar{x}(p)=\sqrt{1-p}$, which is independent of the initial state and strictly decreasing (red dashed line).

\begin{figure}[t]
\begin{center}
\includegraphics[scale=0.5]
{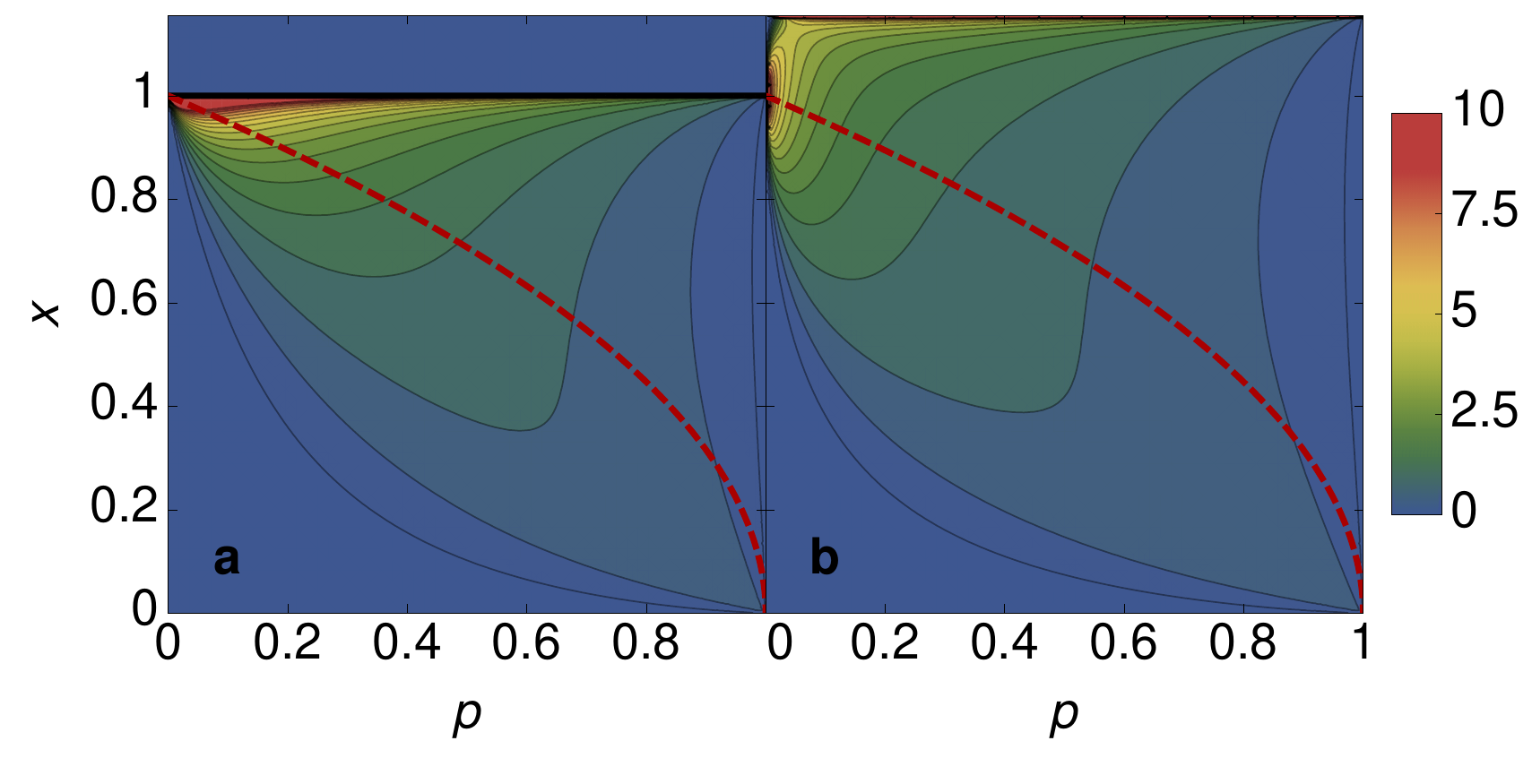}  
\caption{\textbf{Probability distribution of entanglement.} Probability distribution of scaled multipartite entanglement in the CS over the ensemble of experimental outcomes $\{a\}$ of a non-demolition bath for different initial states: \textbf{a,} $\rho_{11}=1/2$, \textbf{b,} $\rho_{11}=1/4$. The figure is the same for any G-invariant measure of a system of qudits if only a single qubit is coupled to a non-demolition bath. The black continuous line shows the maximum possible value of entanglement $x^{\text{max}}(p)$ while the red dashed line corresponds to the mean entanglement $\bar{x}(p)$.}  
\label{fig:fig2}
\end{center}
\end{figure}

A quite different picture arises when the channel corresponds to a non-demolition bath. Thus $\hat{H}_{{\rm CS}}^{\rm (ch)}\propto \hat{J}=\frac{\Delta}{2}\hat{\sigma}_{z}=\hat{O}(a,s,s')$ \cite{qt1}, $\hat{\sigma}_{z}$ is the third of the Pauli's matrices, and $\Delta^2$ the ``phase-flip" rate.  In this so-called dephasing channel, the distribution $P_{G}(x,t)$ is independent of $\rho_{10}$ and has a sharp, time independent boundary given by $x^{\rm max}=\left(4\rho_{11}\rho_{00}\right)^{-1/2}$. For maximally mixed  $\hat{\rho}_{\rm ch}(0)$ the upper limit is $x^{\rm max}=1$, while $x^{\rm max}$ has square-root divergences for pure initial states $\rho_{11}=1,0$. In Fig.~\ref{fig:fig2} we plot $P_{G}(x,p)$ for the Markovian case with $I(\omega)=\pi^{-1}$ and rescaled time $p(t)=1-{\rm e}^{-\Delta^{2}t}$.

\begin{figure}[t]
\begin{center}
\includegraphics[scale=0.49]
{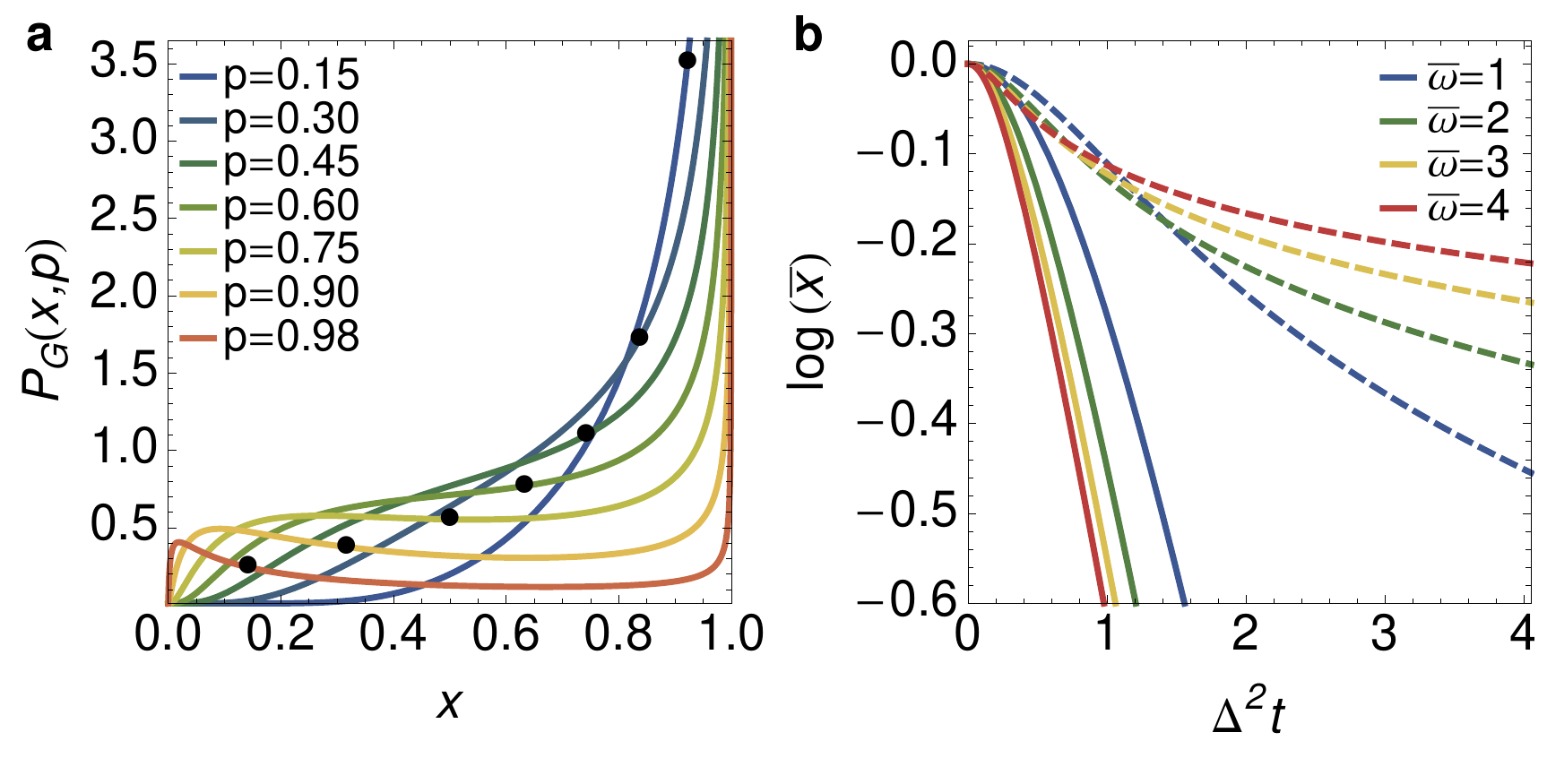}
\caption{\textbf{Evolution of entanglement distribution and mean entanglement.} \textbf{a,} Probability distribution of multipartite entanglement at different values of the scaled time $p$. The black dots mark the mean entanglement $\bar{x}(p)$. \textbf{b,} Logarithm of the average entanglement $\bar{x}$ as a function of $\Delta^{2} t$ for ohmic (continuous lines) and superohmic (dashed lines) environments and different values of the effective Debye frequency ($\bar{\omega}=\omega_{d}/ \Delta^{2}$). Both behaviors demonstrate a transition from exponential to much slower algebraic decay, also characteristic of the approach $p(t \to \infty) \to 1$.}     
\label{fig:fig3}
\end{center}
\end{figure}

For this class of environments the maximum of $P_{G}(x,t)$, which signals the most likely values of the multipartite entanglement, do not lie near the average value $\bar{x}(t)$. Instead, the distributions peaks around the boundary $x=x^{\text{max}}$, where it has an integrable divergence (See Fig.~\ref{fig:fig2}). This is clearly seen in the Markovian case with $\rho_{11}=1/2$, where
\begin{equation*}
\begin{split}
P_{G}(x,p)=&\sqrt{\frac{2(1-p)}{\pi|\log(1-p)|}}\frac{1}{x \sqrt{1-x^2}} \\
& \times \exp\left[{-\frac{\log^2\left(\frac{1+\sqrt{1-x^{2}}}{x}\right)}{2|\log(1-p)|}}\right]\,,
\end{split}
\end{equation*} 
and that we show in Fig.~\ref{fig:fig3}. When time increases a second maximum appears near (but not exactly at) $x=0$, showing that the average $\bar{x}(p)$ (red dashed line in Fig.~\ref{fig:fig2} and points in Fig.~\ref{fig:fig3}) misses the most likely values of multipartite entanglement. 

In this case, the dependence of our results on the memory scale of the environment is easily checked. For this we introduce non-Markovian effects using  a spectral density $I(w) \ne \pi^{-1}$ \cite{open1} (see Supplementary Information), and find that this amounts only to a redefinition of time $t \to q(t)$ in the purely ohmic result by means of
\begin{equation*}
q(t)=\frac{t^{2}}{2}\int_{0}^{\infty}I(\omega)\left(\frac{\sin{\omega t/2}}{\omega t/2}\right)^{2}d\omega \,.
\end{equation*}
The parameterization $p(t)=1-{\rm e}^{-\Delta^{2}q(t)}$ for the time is of most convenience in this case, leading to the mean entanglement $\bar{x}(p)=\sqrt{1-p}$. The function $p(t)$ rules the actual speed at which the distribution and its first moment attain their asymptotic form. It is given for exponential cut-off at Debye frequency $\omega_{d}$ by
\begin{equation*}
p(t\gg \omega_{d}^{-1}) = \begin{cases}
1-(1+\omega_{d}^{2}t^{2})^{1/\pi \bar{\omega}}{\rm e}~^{-\Delta^{2}t} &  \text{(ohmic)}\,, \\
1-1/(1+\omega_{d}^{2}t^{2})^{1/\pi \bar{\omega}} &  \text{(superohmic)}\,,
\end{cases}
\end{equation*}
where we introduced $\bar{\omega}=\omega_d / \Delta^{2}$. The above relation implies that the decay of the averaged entanglement is only algebraic for superohmic environments (see Fig.~\ref{fig:fig3}). Remarkably, however, the approach to the limit $\omega_{d}t \to \infty$ is such that for dephasing channels the average entanglement is never zero for finite times.

\subsection{Finite disentanglement time}

As discussed above, the most likely values of mutipartite entanglement in conditional states are in general not related to their average value. A prime exception takes place when there exists a finite time $t=\tau$ such that $\bar{x}(\tau)=0$, since it entails that at $\tau$ all entanglement measures of the multipartite conditional states vanish identically over the whole ensemble of experimental outcomes. The conditions for this event to occur as well as the dependence of the disentanglement time $\tau$ with the parameters of the system are issues of obvious practical interest. Here we answer these questions for a large class of systems where at least one channel $k=l$ satisfies 
\begin{equation}
\label{eq:cond}
\hat{O}_{l}(a_{l},s,s') = \hat{O}_{l}u_{l}(s,s')
\end{equation}
for some function $u_{l}(s,s')$, and operator $\hat{O}_{l}$. This class of systems includes all Markovian and all non-demolition baths, the Jaynes-Cummings model and the damped harmonic oscillator among others \cite{qt1}.

We notice two immediate consequences of the form of equation~\eqref{eq:barx} for the mean entanglement: First, if (\ref{eq:cond}) holds for all channels, the mean entanglement reduces to
\begin{equation}
\label{eq:xbargen}
\bar{x}(t)={\rm e}^{-\sum_{k}\frac{2 \Re}{d^{(k)}}{\rm Tr}\left[\hat{J}_{k}^{\dagger}\hat{{\cal O}}_{k}\right]\int_{0}^{t}\int_{0}^{s}\alpha_{k}(s-s')u_{k}(s,s')ds ds'}.  
\end{equation} 
Second, in the Markovian case, $u(s,s')=1$, $\hat{O}=\hat{J}$, $\bar{x}(t)$ is independent of the internal dynamics and strictly exponential; no finite $\tau$ exists and the mean entanglement disappears only asymptotically in time.

\begin{figure}[t]
\begin{center}
\includegraphics[scale=0.48]
{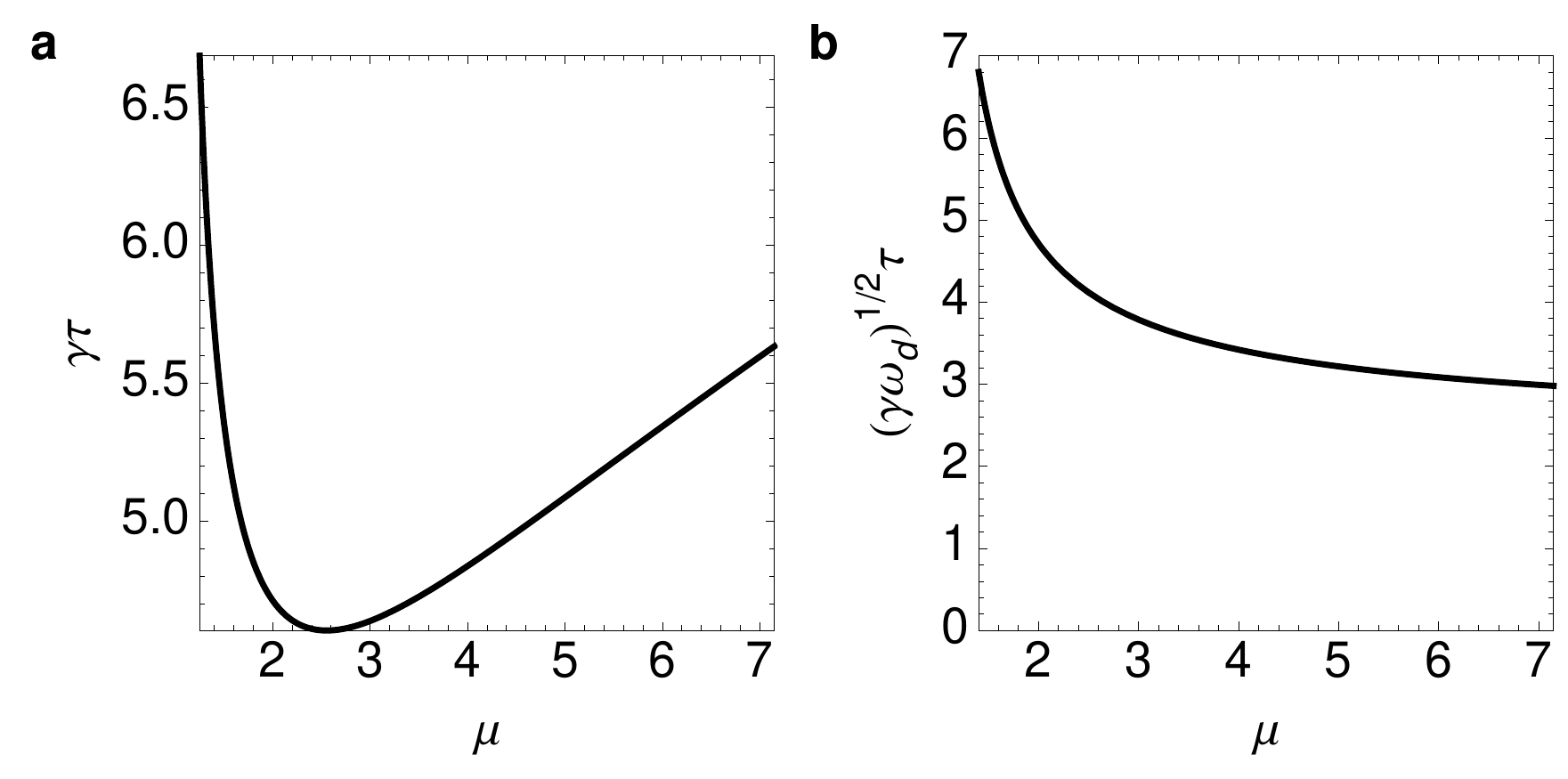}
\caption{\textbf{Disentanglement times.} \textbf{a,} Dependence of the disentanglement times $\tau$ with the effective coupling $\mu=2\gamma /\omega_{d}$ for an arbitrary multipartite system where a qubit without internal dynamics is coupled to a zero-temperature bath with a characteristic memory scale $\omega_{d}^{-1}$. \textbf{b,} Same as in \textbf{a} showing that for large coupling the natural scale for the disentanglement time $\tau$ is given by $(\gamma\omega_{d})^{-1/2}$.}      
\label{fig:fig4}
\end{center}
\end{figure}

More generally, the product form of both $P(a,0)$ and $f(a,t)$ implies that the vanishing of the factor associated with a single channel will automatically disentangle the whole system. Let us focus on this particular channel and assume that it couples a qubit with frozen internal dynamics to a non-Markovian zero-temperature bath with memory kernel $\alpha(s)=(\omega_{d}/2){\rm e}^{-\omega_{d}|s|}$. The exact dynamics of the total system for arbitrary $\alpha(s)$ is obtained with the choice $\hat{O}=\sqrt{\gamma}\hat{\sigma}_{-}$ as long as the function $u(s,s')$ satisfies $u(s,s')=c(s)/c(s')$ and $dc(s)/ds=-\gamma \int_{0}^{s}\alpha(s-s')c(s')ds'$, which can be solved for the case we are considering (see Supplementary Information). After substituting our result into \eqref{eq:xbargen} we obtain
\begin{equation*}
\label{eq:fqb}
\bar{x}(t) \propto {\rm exp}\left\{-\gamma \int_{0}^{t}\frac{\frac{\omega_{d}}{2\Omega}\sinh\Omega s}{\frac{\omega_{d}}{2\Omega}\sinh\Omega s +\cosh \Omega s}ds\right\}\,,
\end{equation*}
where $\Omega=(\omega_{d}/2)\sqrt{1-\mu}$ and $\mu=2\gamma/w_{d}$. We notice that for weak and strong coupling ($\mu \le 1$) the average entanglement is not zero for all finite times. In contrast, in the ultrastrong coupling regime ($\mu >1$) we get $\bar{x}(\tau)=0$ for 
\begin{equation*}
\label{eq:tau}
\gamma \tau=\frac{\mu}{\sqrt{\mu-1}}\left(\pi-\arctan\sqrt{\mu-1}\right),
\end{equation*}
rigorously showing the existence of finite disentanglement times in all conditional states of arbitrary multipartite systems where at least one single qubit is coupled with a non-Markovian bath. Remarkably, the disentanglement time increases with $\mu$ for strong enough coupling, as shown in Fig.~\ref{fig:fig4}.

Our last considerations are directly extended to the unconditional dynamics of the open system (entanglement sudden death \cite{ESD1,ESD2,ESD3,ESD4,ESD5,ESD6}). The mixed state describing the unconditional evolution is given in complete generality by an average of the pure conditional states \cite{qt0,qt1,qt2}
\begin{equation*}
\hat{\rho}^{\rm unc}(t)=\int P(a,t)|\phi(a,t)\rangle \langle \phi(a,t)| \,da.
\end{equation*}
Therefore, the set \{$|\phi(a,t)\rangle, P(a,t)da\}$ constitutes a valid convex decomposition of $\hat{\rho}^{\rm unc}(t)$, and since the amount of entanglement $G(\hat{\rho}^{\rm unc}(t))$ is defined by a convex roof construction, we have
\begin{equation*}
\label{eq:uncond}
\frac{G(\hat{\rho}^{\rm unc}(t))}{ G(\hat{\rho}^{\rm unc}(0)) } \le \bar{x}(t).
\end{equation*}
Therefore all measures of multipartite entanglement of an open system interacting with an arbitrary photonic environment must be zero at a finite time if at least one qubit is strongly coupled with a non-Markovian bath at zero temperature. The important extension of this result to finite-temperature baths is presently under study \cite{us}.

\subsection{Tomography of entanglement} 
  
To conclude, in this last section we show how scaling law \eqref{eq:resI} provides us with a geometrical and operational interpretation of the amount of entanglement in conditional states of multipartite systems. This follows from the observation that, associated with the pure total state $|\Psi(t)\rangle$, the unconditional density matrix describing the environment $\hat{\rho}_{\rm E}(t)={\rm Tr}_{\rm CS} |\Psi(t)\rangle \langle \Psi(t)|$ \cite{qt3} has the following representation in coherent states  
\begin{equation*}
\label{eq:Q}
Q(a,t)={\cal Z}^{-1}\langle a|\left[{\rm Tr}_{\rm CS} |\Psi(t)\rangle \langle \Psi(t)|\right]|a\rangle,
\end{equation*}
where $Q(a,t)$ is the so-called Husimi function of the bath \cite{hus}. Using equation~(\ref{eq:Psizeta}) then leads to\begin{equation*}
Q(a,t)=P(a,t),
\end{equation*}
and therefore equation~(\ref{eq:resI}) relates the possible values of the entanglement of the conditional state with the morphology of the Husimi function of the bath; a quantity accessible to direct measurement. 

In the following we will consider systems where equation~(\ref{eq:cond}) holds. In this case we easily obtain
\begin{equation}
\label{eq:idmark}
\frac{x(a,t)}{x(a',t)}={\rm e}^{-\left(N_{a}-N_{a'}\right)}\frac{Q(a',t)}{Q(a,t)},
\end{equation}
where $N_{a}=|a|^{2}$ is the mean number of photons in the state $|a\rangle$. Hence, the numerical value of the entanglement of the conditional state after a {\it single} measurement with experimental outcome $\{ a'\}$ together with the unconditional Husimi function of the bath $Q(a',t)$  automatically predicts {\it all} the entanglement measures of {\it all} possible multipartite conditional states. {\it Quantum tomography of the bath is actually measuring entanglement of the conditional states}. 

Finally, known upper bounds for the Husimi function \cite{hus} can be used to provide a {\it lower} bound to entanglement
\begin{equation*}
x(a,t)\ge{\rm e}^{-N_{a}}\bar{x}(t),
\end{equation*}  
which is particularly useful in the experimentally accessible case where $|a\rangle$ lies close to the ground state. 

\subsection{Summary}

The derivation of a scaling law relating the amount of entanglement in the state of a multipartite system just after its photonic environment is measured, to the probability distribution of the experimental outcomes, constitutes the central result of this paper. It indicates that for bosonic environments, multipartite entanglement is severely constrained by indirect measurement. Specifically, we use equation (\ref{eq:resI}) to demonstrate that all possible normalized measures of multipartite entanglement in the conditional state have the same distribution over the ensemble of measurement outcomes of the bath. We construct the full distribution in standard cases with one open channel and show that for non-demolition environments non-Markovian effects appear only through a redefinition of the time variable. We rigorously prove disentanglement in both pure conditional and mixed unconditional states of multipartite systems where one qubit is coupled with a zero temperature non-Markovian bath, with a counter intuitive dependence of the disentanglement times on the coupling strength. Finally, we showed that due to the scaling, for a large class of systems, multipartite entanglement of the conditional states is directly measured by quantum tomography of the bath through the Husimi function.

\subsection{Methods}

{\bf Scaling function}. Here we calculate explicitly the scaling function in equation \eqref{eq:resI}. By definition, $f(a,t)$ must satisfy
\begin{equation*}
G(\hat{\cal U}(a,t)|\phi(0)\rangle)=f(a,t)G(|\phi(0)\rangle)
\end{equation*}
with
\begin{equation*}
\hat{\cal U}(a,t) ={\cal T}{\rm e}^{\int_{0}^{t}\left[-\frac{i}{\hbar}\hat{H}_{\rm CS}+\sum_{k}\left(z^{*}_{k}(s)\hat{J}_{k}-\hat{J}^{\dagger}_{k}\hat{{\cal O}}_{k}(s)\right)\right]ds}.
\end{equation*}
Assuming that the Hamiltonian $\hat{H}_{\rm CS}$, the Linblad operators $\hat{J}_{k}$, and the convolutionless operators $\hat{{\cal O}}_{k}(s)$ act locally on the subsystems, the operator $ \hat{\cal U}$ admits the factorization
\begin{equation*}
\hat{\cal U}(a,t) =\prod_{i=1}^{N}\hat{\cal U}^{(i)}(a^{(i)},t).
\end{equation*}
Here
\begin{equation*}
\hat{\cal U}^{(i)}(a^{(i)},t) ={\cal T}{\rm e}^{\int_{0}^{t}\left[-\frac{i}{\hbar}\hat{H}_{\rm CS}^{(i)}+\sum_{k^{(i)}}\left(z^{*}_{k^{(i)}}(s)\hat{J}_{k^{(i)}}-\hat{J}^{\dagger}_{k^{(i)}}\hat{{\cal O}}_{k^{(i)}}(s)\right)\right]ds}
\end{equation*}
is a local operators acting on the $i$th subsystem, and $\{k^{(i)}\}$ denotes the indexes of the Lindblad operators acting on the same subsystem. We now introduce operators
\begin{equation*}
\hat{U}^{(i)}(a^{(i)},t)=|{\rm det~}\hat{\cal U}^{(i)}(a^{(i)},t)|^{-1/d^{(i)}}\hat{\cal U}^{(i)}(a^{(i)},t)
\end{equation*}
with $|{\rm det~}\hat{U}^{(i)}|=1$, since $d^{(i)}$ is the dimension of the Hilbert space of the $i$th subsystem, and use the homogeneity of the function $G$ to obtain
\begin{equation*}
\begin{split}
G(\hat{\cal U}(a,t)|\phi(0)\rangle)=& \prod_{i=1}^{N}\left|{\rm det~}\hat{\cal U}^{(i)}(a^{(i)},t)\right|^{2/d^{(i)}} \\
&\times G\left(\left[\prod_{i=1}^{N}\hat{U}^{(i)}(a^{(i)},t)\right]|\phi(0)\rangle\right).
\end{split}
\end{equation*}
Since by definition $G(|\phi\rangle)$ is invariant under local transformations with determinant one, we can identify 
\begin{equation*}
f(a,t)=\prod_{i=1}^{N}\left|{\rm det~}\hat{\cal U}^{(i)}(a^{(i)},t)\right|^{2/d^{(i)}}.
\end{equation*}
To proceed we focus on the individual terms in the product. Under mild conditions, the determinant of $\hat{\cal U}^{(i)}$ admits the representation
\begin{equation*}
{\rm det~}\hat{\cal U}^{(i)}(a^{(i)},s)={\rm e}^{{\rm Tr~log~}\hat{\cal U}^{(i)}(a^{(i)},s)},
\end{equation*}
and therefore
\begin{equation*}
{\rm log~}\left|{\rm det~}\hat{\cal U}^{(i)}(a^{(i)},s)\right|^{2/d^{(i)}}=\frac{2 \Re }{d^{(i)}}{\rm Tr~log~}\hat{\cal U}^{(i)}(a^{(i)},s).
\end{equation*}
To evaluate this last expression we start from the equation of motion for $\hat{\cal U}^{(i)}$,
\begin{eqnarray}
&&\frac{\partial}{\partial s}\hat{\cal U}^{(i)}(a^{(i)},s)= -\frac{i}{\hbar}\hat{H}_{\rm CS}^{(i)}\hat{\cal U}^{(i)}(a^{(i)},s)\nonumber \\ &&+\left[\sum_{k^{(i)}}\left(z^{*}_{k^{(i)}}(s)\hat{J}_{k^{(i)}}-\hat{J}^{\dagger}_{k^{(i)}}\hat{{\cal O}}_{k^{(i)}}(s)\right)\right]\hat{\cal U}^{(i)}(a^{(i)},s), \nonumber
\end{eqnarray}
to write
\begin{eqnarray}
&&\frac{\partial}{\partial s}{\rm log~}\left|{\rm det~}\hat{\cal U}^{(i)}(a^{(i)},s)\right|^{2/d^{(i)}}=\frac{2 }{d^{(i)}} \times \nonumber \\ && \Re {\rm Tr~}\left[-\frac{i}{\hbar}\hat{H}_{\rm CS}^{(i)}+\sum_{k^{(i)}}\left(z^{*}_{k^{(i)}}(s)\hat{J}_{k^{(i)}}-\hat{J}^{\dagger}_{k^{(i)}}\hat{{\cal O}}_{k^{(i)}}(s)\right)\right], \nonumber
\end{eqnarray}
where we used the cyclic property of the trace. Further simplifications take place after noticing that $\hat{H}_{\rm CS}$ is hermitian and that by construction $\hat{J}_{k}$ is traceless, then  
\begin{equation*}
\frac{\partial}{\partial s}{\rm log~}\left|{\rm det~}\hat{\cal U}^{(i)}(a^{(i)},s)\right|^{\frac{2}{d^{(i)}}}=-\frac{2 \Re}{d^{(i)}}{\rm Tr~}\sum_{k^{(i)}}\hat{J}^{\dagger}_{k^{(i)}}\hat{{\cal O}}_{k^{(i)}}(s),
\end{equation*}
which is trivially integrated. Substituting the result in $f(a,t)$ and collecting the sums in the exponents via  $\sum_{i}\sum_{k^{(i)}}=\sum_{k}$ we finally get 
\begin{equation*}
f(a,t)={\rm e}^{-\sum_{k}\frac{2 \Re }{d^{(k)}}\int_{0}^{t}{\rm Tr}\left[\hat{J}_{k}^{\dagger}\hat{{\cal O}}_{k}(s)\right]ds},
\end{equation*}
thus completing the construction of the scaling function.

{\it Acknowledgements}. We kindly acknowledge illuminating discussions with Christopher Eltschka and Jens Siewert. C.V. is thankful for the hospitality extended to him by Institut f\"ur Theoretische Physik at Universit\"at Regensburg. J.~D.~U. acknowledges financial support from the Deutsche Forshungsgemeinschaft through  Forschergruppe 760: Scattering Systems with Complex Dynamics. C.~V. and J.~D.U acknowledge Klaus Richter for his support to this project.

\newpage
\rule{5pt}{0pt}
\newpage

\section{Supplementary Information}

\subsection{Open system Hamiltonian}
In this section we derive within the usual system-plus-bath approach (see Refs.~4,~24,~25 and 26) the Hamiltonian of an open system appearing in equation (1) in the main text. For this, we consider the system dynamics in a convenient interaction picture and show how to redefine the Lindblad operators to make them traceless. For simplicity we consider a single channel for the derivation; the generalization to arbitrary number of channels is straightforward and will be presented at the end of the section.

Our starting point is the total Hamiltonian $\hat{H}$ describing a central system (CS) linearly coupled with a bosonic bath (B),
\begin{equation}
\label{eq:Htot}
\hat{H}=\hat{H}_{\rm CS}+\hat{H}_{\rm CSB} +\hat{H}_{\rm B}
\end{equation}
with environment and coupling terms given by
\begin{align*}
\label{eq:Hs}
\hat{H}_{\rm B}&=\sum_{\lambda}w_{\lambda}\hat{a}^{\dagger}_{\lambda}\hat{a}_{\lambda}\,, \\
\hat{H}_{\rm CSB}&=i\sum_{\lambda}g_{\lambda}\left(\hat{L}\hat{a}^{\dagger}_{\lambda}-\hat{L}^{\dagger}\hat{a}_{\lambda}\right).
\end{align*}
We consider a bosonic bath with frequency $\omega_\lambda$ in its $\lambda$th mode, for which $a_\lambda$ and $a_\lambda^\dagger$  correspond to anihilation and creation operators, respectively. The operator $\hat{L}$ acts on the CS and is usually dubbed Lindblad operator. Together with the real coupling constants $g_{\lambda}$ it describes the interaction of the central system with all the modes of the bath. The total Hilbert space ${\cal H}={\cal H}_{\rm CS} \otimes {\cal H}_{\rm B}$ of the system is spanned by product states $|e_{i}\rangle \otimes |a_{1},\ldots,a_{\lambda},\ldots\rangle$, where $\{|e_{i}\rangle\}$ is any basis of the CS Hilbert space, and $|a\rangle=|a_{1},\ldots,a_{\lambda},\ldots\rangle $ is a normalized coherent state of the environment, i.e., a common eigenstate of the annihilation operators of the bath, $\hat{a}_{\lambda}|a\rangle = a_{\lambda}|a\rangle$ with $\langle a| a\rangle=1$.

Introduce now the transformations 
\begin{align*}
\hat{a}_{\lambda}&=\hat{b}_{\lambda}-i\frac{g_\lambda}{\omega_\lambda} \alpha \,, \\
\hat{L}&=\hat{J}+\alpha,
\end{align*}
with complex constant $\alpha$. Observe that the coherent states $|a\rangle$ are also eigenstates of $\hat{b}$. Then, up to a physically irrelevant constant term, Hamiltonian \eqref{eq:Htot} becomes
\begin{equation*}
\hat{H}=\hat{H}_{\rm CS}-\sum_{\lambda}\frac{g_{\lambda}^{2}}{w_{\lambda}}\left(\alpha^{*}\hat{J}+\alpha\hat{J}^{\dagger}\right)+\sum_{\lambda}w_{\lambda}\hat{b}_{\lambda}^{\dagger}\hat{b}_{\lambda}\,.
\end{equation*}
In order to obtain a traceless coupling operator $\hat{J}$, we take advantage of the freedom introduced by $\alpha$ and choose $\alpha=d^{-1}{\rm Tr~}\hat{L}$, where $d$ is the dimension of the Hilbert space of the CS. Thus the hermitian redefinition of the CS Hamiltonian
\begin{equation}
\label{eq:ren} 
\hat{H}_{\rm CS} \to \hat{H}_{\rm CS}-\sum_{\lambda}\frac{g_{\lambda}^{2}}{w_{\lambda}}\left(\alpha^{*}\hat{J}+\alpha\hat{J}^{\dagger}\right)
\end{equation}
allow us to work with traceless Lindblad operators in full generality. Since this change, up to a constant, is the only consequence of the transformation, we stick to the initial notation and write the total Hamiltonian as in equation (\ref{eq:Htot}) but with $\hat{J}$ instead of $\hat{L}$, and keep in mind renormalization \eqref{eq:ren} for the CS Hamiltonian.

The interaction picture for the total time-evolution operator $\hat{U}(t)={\rm e}^{-\frac{i}{\hbar}\hat{H}t}$ is defined via the unitary transformation
\begin{equation*}
\hat{U}_{\rm int}(t)={\rm e}^{\frac{i}{\hbar}\hat{H}_{\rm B}t}\hat{U}(t){\rm e}^{-\frac{i}{\hbar}\hat{H}_{\rm B}t} \,,
\end{equation*}
and absorbs the explicit dependence on the Hamiltonian of the bath into a redefinition of the bath operators $ \hat{a}_{\lambda}^{\rm int}(t)=\hat{a}_{\lambda}{\rm e}^{-iw_{\lambda}t}$, rendering the transformed Hamiltonian explicitly time-dependent. Formal solution of the Schr\"odinger equation in the interaction picture yields  
\begin{equation*}
\hat{U}_{\rm int}(t)={\cal T}{\rm e}^{-\frac{i}{\hbar}\int_{0}^{t} (\hat{H}_{CS}(s)+\hat{H}_{I}(s))ds}\,,
\end{equation*}
where $\hat{H}_{CS}(s)$ is the renormalized CS Hamiltonian, ${\cal T}$ is the time-ordering operator, and
\begin{equation*}
\hat{H}_{I}(s)=i\sum_{k,\lambda}g_{k,\lambda}\left(\hat{J}_{k}\hat{a}_{k,\lambda}^{\dagger}{\rm e}^{i \omega_{k,\lambda}s}-\hat{a}_{k,\lambda}\hat{J}_{k}^{\dagger}{\rm e}^{-i \omega_{k,\lambda}s}\right)
\end{equation*} 
is the interaction Hamiltonian. In this last step we already introduced the result for the general case with an arbitrary number of channels indexed by $k$.

\subsection{Entanglement distributions}

Here we evaluate the probability distributions of multipartite entanglement for conditional states in the CS over the ensemble of experimental outcomes of measurements on the environments studied in the main text. Using the scale law stated in equation (6) of the main text, the probability distribution of entanglement, as given by equation (7) of the main text, takes the manifestly universal form
\begin{equation*}
\begin{split}
P_{G}(x,t) &= \frac{1}{\mathcal{Z}} \int\frac{f(a,t) \, {\rm e}^{-|a|^{2}}}{x} \\
& \quad \times \delta\left(x-\frac{f(a,t)}{{\rm Tr~}\hat{\rho}_{{\rm CS}}(0)\hat{\cal U}(a,t)^{\dagger} \hat{\cal U}(a,t)}\right) da \,,
\end{split}
\end{equation*}
where we used equations (3) and (5) from the main text for simplifications. The particular dependence of the scaling function $f(a,t)$ and $\hat{\cal U}(a,t)^{\dagger} \hat{\cal U}(a,t)$ on the bath measurement outcomes $\{a\}$ must be found in each particular case of interest from their definitions (see Methods section in the main text):
\begin{align*}
f(a,t)&={\rm e}^{-\sum_{k}\frac{2\Re}{d^{(k)}}\int_{0}^{t}\int_{0}^{s}\alpha_{k}(s-s'){\rm Tr}\left[\hat{J}_{k}^{\dagger}\hat{O}_{k}(a_{k},s,s')\right]ds ds'}\,, \\
\hat{\cal U}(a,t) &={\cal T}{\rm e}^{\int_{0}^{t}\left[-\frac{i}{\hbar}\hat{H}_{\rm CS}+\sum_{k}\left(z^{*}_{k}(s)\hat{J}_{k}-\hat{J}^{\dagger}_{k}\hat{{\cal O}}_{k}(s)\right)\right]ds}.
\end{align*}

Further simplifications in the expression for the entanglement distribution are obtained if we assume that only a single subsystem is coupled to a channel and use the locality of $\hat{H}_{\text{CS}}$. In this case, the part of $ {\rm Tr~}\hat{\rho}_{{\rm CS}}(0)\hat{\cal U}(a,t)^{\dagger} \hat{\cal U}(a,t)$ which traces over the subsystems not coupled with the open channel can be calculated, leaving only the reduced density matrix describing the initial state of the coupled subsystem (denoted by ``ch"). If in addition we suppose that
\begin{equation*}
\hat{O}^{{\rm (ch)}}(a_{{\rm (ch)}},s,s') = \hat{O}^{{\rm (ch)}}u^{{\rm (ch)}}(s,s'),
\end{equation*}
for some operator $\hat{O}$ and function $u(s,s')$, then $f(a,t)=f(t)$ is independent of $\{a\}$, and all variables $a_{(k \ne {\rm ch})}$ in the probability distribution can be integrated out by performing the Gaussian integral left for them. All together, the key technical difficulties we must face to obtain the entanglement distribution are the evaluation of the function
\begin{equation*}
F(a_{\rm (ch)},t)={\rm Tr}_{\rm ch}~\hat{\rho}_{{\rm ch}}(0)\hat{\cal U}^{\rm (ch)}(a_{\rm (ch)},t)^{\dagger} \hat{\cal U}^{\rm (ch)}(a_{\rm (ch)},t),
\end{equation*}
and the calculation of the integral
\begin{equation}
\label{eq:Pg}
P_{G}(x,t)=\frac{f(t)}{x}\int\frac{{\rm e}^{-|a_{\rm (ch)}|^{2}}}{{\cal Z}_{\rm (ch)}}\delta\left(x-\frac{f(t)}{F(a_{\rm (ch)},t)}\right) da_{\rm (ch)}. 
\end{equation}
We now follow this program for the two physically relevant situations with one open channel considered in the main text.

\subsubsection{Markovian channel at zero temperature}

For a Markovian channel at zero temperature coupled to a qubit, $\hat{J}=\sqrt{\gamma}\hat{\sigma}_{-}$ with $\sigma_-$ the deexcitation operator and $\gamma$ the decay rate, $\hat{O}(a,s,s')=\hat{J}$ (see Refs.~24,~25 and 26), and $f(a,t)={\rm e}^{-\gamma t/2}$. 

Taking for simplicity a frozen internal dynamics $\hat{H}_{\rm CS}^{\rm (ch)}=0$, the equation of motion for the non-unitary propagator is
\begin{equation*}
\label{eq:eqmu}
\frac{\partial}{\partial s}\hat{\cal U}(a,s)=\left[\sqrt{\gamma}z^{*}(a,s)\hat{\sigma}_{-}-\frac{\gamma}{2}\hat{n}_{-}\right]\hat{\cal U}(a,s)
\end{equation*} 
with initial condition $\hat{\cal U}(a,0)=\hat{I}$. Here $\hat{n}_{-}=\hat{\sigma}_{+} \hat{\sigma}_{-}$, and $ z(s)=\sum_{\lambda}g_{\lambda}a_{\lambda}{\rm e}^{-i \omega_{\lambda}s}$ is a sum over the modes of the bath defining the open channel. The above equation can be solved by introducing the interaction picture in $\hat{n}_{-}$ and exploiting the commutator $[\hat{n}_{-},\hat{\sigma}_{\pm}]=\pm\hat{\sigma}_{\pm}$, the result is
\begin{equation*}
 \hat{\cal U}(a,t)={\rm e}^{-\frac{\gamma}{2}\hat{n}_{-}t}{\rm e}^{\sqrt{\gamma}\int_{0}^{t}z^{*}(a,s){\rm e}^{-\frac{\gamma}{2}s}ds \hat{\sigma}_{-}}.
\end{equation*}
Explicit calculation using the algebraic properties of the Pauli matrices then gives
\begin{equation*}
\begin{split}
\hat{\cal U}(a,t)^{\dagger}\hat{\cal U}(a,t) &=\hat{I}+\left({\rm e}^{-\gamma t}-1+|\vec{a}^{\tau}\vec{g}(t)|^{2}\right)\hat{n}_{-} \\ 
&\quad +\left[\vec{a}^{\tau}\vec{g}(t)\hat{\sigma}_{+} + (\vec{a}^{\tau}\vec{g}(t))^* \hat{\sigma}_{-}\right], 
\end{split}
\end{equation*}
where 
\begin{align*}
\vec{a}^{\tau}\vec{g}(t)&=\sqrt{\gamma}\int_{0}^{t}z^{*}(a,s){\rm e}^{-\frac{\gamma}{2}s}ds  \\
&= i \sqrt{\gamma}\sum_{\lambda}a_{\lambda}g_{\lambda}\frac{{\rm e}^{-i\left(\omega_{\lambda}-i\frac{\gamma}{2}\right)t}-1}{\omega_{\lambda}-i\frac{\gamma}{2}}.
\end{align*}
Consequently, $F(a,t)$ depends on the integration variables $a_{\lambda}$ only through the linear combination $y(a,t)=\vec{a}^{\tau}\vec{g}(t)$. The distribution function for $y=y(a,t)$ is readily calculated,
\begin{align*}
P_{y}(y,t) &= \int\!\frac{{\rm e}^{-|a|^{2}}}{{\cal Z}}\delta\left(y-\vec{a}^{\tau}\vec{g}(t)\right)da \\
&=\frac{1}{\pi |\vec{g}(t)|^2} \exp\left(-\frac{|y|^2}{|\vec{g}(t)|^2}\right) ,
\end{align*}
yielding a Gaussian function with width given by
\begin{equation*} 
|\vec{g}(t)|^{2}=\gamma \sum_{\lambda}\frac{g_{\lambda}^{2}}{\omega_{\lambda}^{2}+(\gamma/2)^{2}}\left({\rm e}^{-\gamma t}-2{\rm e}^{-\frac{\gamma}{2}t}\cos \omega_{\lambda}t +1\right),
\end{equation*}
which can be express in terms of the spectral density of the bath. For the Ohmic case we get
\begin{align*} 
p(t):=|\vec{g}(t)|^{2}&=\frac{\gamma}{2 \pi} \int_{-\infty}^{\infty}\frac{{\rm e}^{-\gamma t}-2{\rm e}^{-\frac{\gamma}{2}t}\cos \omega t +1}{\omega^{2}+(\gamma/2)^{2}}d\omega \nonumber \\
&=1-{\rm e}^{-\gamma t}.
\end{align*}

After collecting our results, the probability distribution of entanglement \eqref{eq:Pg} can be recast into
\begin{equation*}
P_{G}(x,p)=\frac{{\rm e}^{-\frac{\gamma}{2}t}}{x}\int\frac{{\rm e}^{-\frac{|y|^{2}}{p(t)}}}{\pi p(t)}\delta\left(x-\frac{{\rm e}^{-\frac{\gamma}{2}t}}{\tilde{F}(y)}\right)dy
\end{equation*}
with
\begin{equation*}
\tilde{F}(y)=\Re {\rm Tr~}\hat{\rho}_{{\rm ch}}(0)\left[\hat{I}+\left({\rm e}^{-\gamma t}-1+|y|^{2}\right)\hat{n}_{-}+y\hat{\sigma}_{+}\right].
\end{equation*}
This last integral can be evaluated to finally obtain
\begin{widetext}
\begin{align*}
P_{G}(x,t) &= 2\frac{1-p(t)}{x^{3}\rho_{11}p(t)} \\
& \times \exp\left[-\frac{1}{p(t)}\left( \frac{\sqrt{1-p(t)}}{\rho_{11} x} - \frac{1 - \rho_{11}p(t)}{\rho_{11}} + \frac{2|\rho_{10}|^2}{\rho_{11}^2} \right) \right] \text{I}_{0} \! \left[ \frac{2|\rho_{10}|}{\rho_{11}p(t)} \left( \frac{\sqrt{1-p(t)}}{\rho_{11}x} - \frac{1-\rho_{11}p(t)}{\rho_{11}} + \frac{|\rho_{10}|^{2}}{\rho_{11}^{2}} \right)^{1/2} \right], 
\end{align*}
\end{widetext}
where ${\rm I}_{0}(z)$ is the modified Bessel function of zeroth order.

\subsubsection{Non-Markovian dephasing channel}

We now turn to the general dephasing case in $d=d^{({\rm ch})}$ dimensions for which $\hat{O}(a,s,s')=\hat{J}=\hat{H}_{{\rm CS}}^{({\rm ch})}$ (see refs.~25 and 26) and consider an arbitrary spectral density.

Let us first note that in this case
\[
f(a,t)={\rm e}^{-\frac{2}{d} {\rm Tr}\hat{J}^{2}\, \Re \int_{0}^{t}\!\int_{0}^{s}\alpha(s-s')ds'ds}
\]
is independent of $\{a_{\rm (ch)}\}$. Moreover, introducing 
\begin{equation}
\label{eq:qs}
\begin{split}
q(t) :&= \Re \int_{0}^{t}\!\int_{0}^{s}\alpha(s-s')ds'ds \\
&=\frac{t^{2}}{2}\int_{0}^{\infty}\! I(\omega)\left(\frac{\sin{\omega t/2}}{\omega t/2}\right)^{2}d\omega ,
\end{split}
\end{equation} 
we easily get
\begin{equation*}
\hat{\cal U}(a,t)^{\dagger}\hat{\cal U}(a,t)={\rm e}^{-2 \hat{J}^{2}q(t)+2\hat{J}\,\Re\, \vec{a}^{\tau} \vec{g}(t)},
\end{equation*}
with
\begin{equation*}
\vec{a}^{\tau} \vec{g}(t) = i \sum_{\lambda}a_{\lambda}g_{\lambda}\frac{{\rm e}^{-i \omega_{\lambda} t}-1}{\omega_{\lambda}} \,.
\end{equation*}
The probability distribution for the real variable $y(a,t)=\Re\, \vec{a}^{\tau} \vec{g}(t)$ is again given by a Gaussian function, 
\begin{align*}
P_{y}(y,t)&=\int\frac{{\rm e}^{-|a|^{2}}}{{\cal Z}}\delta\left(y-\Re \vec{a}^{\tau}\vec{g}(t)\right)da \nonumber \\ 
&=\frac{1}{\sqrt{\pi |\vec{g}(t)|^{2}}} \exp\left( -\frac{y^{2}}{|\vec{g}(t)|^{2}} \right),
\end{align*}
with width
\begin{equation*}
2 q(t) := |\vec{g}(t)|^{2}=\sum_{\lambda}g_{\lambda}^{2}\left|\frac{{\rm e}^{-i \omega_{\lambda} t}-1}{\omega_{\lambda}}\right|^{2}.
\end{equation*}

The probability distribution of entanglement \eqref{eq:Pg} for this environment simplifies into
\begin{align*}
P_{G}(x,t) &= \frac{{\rm e}^{-\frac{2}{d}q(t){\rm Tr~}\hat{J}^{2}}}{x} \\
& \times \int\frac{{\rm e}^{-\frac{y^{2}}{2q(t)}}}{\sqrt{2\pi q(t)}}\delta\left(x-\frac{{\rm e}^{-\frac{2}{d}q(t){\rm Tr}\hat{J}^{2}}}{\tilde{F}(y)}\right)dy,
\end{align*}
in which 
\begin{equation*}
\tilde{F}(y)={\rm Tr}\hat{\rho}_{{\rm ch}}(0){\rm e}^{2 \hat{J}y-2 \hat{J}^{2}q(t)} .
\end{equation*}

If we now restrict to the case in which the coupled subsystem is a qubit, i.e., $d=2$, the integral can be evaluated to obtain the entanglement distribution
\begin{widetext}
\begin{equation*}
P_{G}(x,t) =\frac{1}{x \sqrt{1-4x^{2}\rho_{11}\rho_{00}}} \sqrt{\frac{1-p(t)}{2\pi |\log(1-p(t))|}} \left\{ \exp \left[ \frac{ \log^{2}\left(\frac{1+\sqrt{1-4x^{2}\rho_{11}\rho_{00}}}{2 x \rho_{11}}\right)}{2\log(1-p(t))} \right] + \exp\left[ \frac{ \log^{2}\left(\frac{1+\sqrt{1-4x^{2}\rho_{11}\rho_{00}}}{2 x \rho_{00}}\right)}{2\log(1-p(t))}\right] \right\} \,,
\end{equation*}
\end{widetext}
where we introduced a convenient re-paremeteriztaion of time $p(t)=1-{\rm e}^{-2{\rm Tr}\hat{J}^{2}q(t)}$.

\subsection{Scaled time for non-Markovian dephasing baths}

The behaviour of the function $q(t)$ for different models of the non-Markovian bath is easily found by selecting the appropriate spectral density. In the continuum limit, this distribution is assumed to have the general form
\begin{equation*}
I(w)=\pi^{-1}(w/w_{d})^{\eta}{\rm e}^{-w/w_{d}}
\end{equation*}
where the cut-off (Debye) frequency $\omega_{d}$ is the maximum frequency of the bath. 

We may then consider different limiting cases. Adopting the standard nomenclature (see Ref.~4), $\eta \to 0, \omega_{d} \to \infty$ defines purely ohmic dissipation, while $\eta \to 0$ defines ohmic dissipation with upper frequency, and $\eta \to 1$ represents superohmic dissipation. Explicit calculation of the integral in Eq.~(\ref{eq:qs}) for the different limiting cases leads to
\begin{widetext}
\begin{equation*}
q(t) = \left\{
\begin{aligned}
&t && \text{(purely ohmic)}\,, \\
& \frac{2t}{\pi} \tan^{-1}w_{d}t-\frac{1}{\pi w_{d}}\log(1+w_{d}^{2}t^{2}) && \text{(ohmic)}\,, \\
& \frac{1}{\pi w_{d}}\log(1+w_{d}^{2}t^{2}) &&  \text{(superohmic)}\,, 
\end{aligned}
\right.
\end{equation*}
\end{widetext}
showing a transition from exponential to algebraic decay for the function $p(t)=1-{\rm e}^{-2{\rm Tr}\hat{J}^{2} q(t)}$.

\subsection{Convolutionless operator $\hat{O}(a,s,s')$ for a non-Markovian bath at zero temperature}

In this last section we evaluate the operator $\hat{O}(a,s,s')$ for a non-Markovian bath at zero temperature, following the techniques of Ref.~25. For a CS with frozen internal dynamics $\hat{H}_{\text{CS}}=0$ and coupled to the bath trough an operator $\hat{J}=\sqrt{\gamma}\sigma_{-}$, indicating amplitude damping, the condition for the existence of the operator  
\begin{equation*}
\hat{O}(a,s,s')=\hat{J}u(s,s')
\end{equation*}
was shown in Ref.~26 of the main text to be that the nonlinear integral equation
\begin{equation*}
\frac{\partial}{\partial s}u(s,s')=\gamma u(s,s')\int_{0}^{s}\alpha(s-s'')u(s,s'')ds''
\end{equation*}
with $u(s,s)=1$ admits a solution. 

Remarkably, this nonlinear problem can be exactly mapped into a linear one by introducing $u(s,s')=c(s')/c(s)$, where
\begin{equation*}
\label{eq:cs1}
\frac{\partial}{\partial s}c(s)=-\gamma \int_{0}^{s}\alpha(s-s')c(s')ds'.
\end{equation*}
If the bath is at zero temperature and has a memory kernel of the Ornstein-Uhlenbleck type
\begin{equation*}
\alpha(s)=\frac{w_{d}}{2}{\rm e}^{-w_{d}|s|},
\end{equation*}
this last problem takes the form
\begin{equation}
\label{eq:cs2}
\frac{\partial}{\partial s}c(s)=-\frac{\gamma w_{d}}{2} \int_{0}^{s}{\rm e}^{-w_{d}(s-s')}c(s')ds',
\end{equation}
which, due to its linearity, can be solved by Laplace transform methods. After some algebra the result is
\begin{equation*}
c(s)=\frac{c(0)}{z_{-}-z_{+}}\left(z_{-}{\rm e}^{z_{+}s}-z_{+}{\rm e}^{z_{-}s}\right)\,,
\end{equation*}
where
\begin{equation*}
z_{\pm} = -\frac{\omega_d}{2} \pm \Omega
\end{equation*}
are the solutions of the associated characteristic equation, and we introduced $\Omega = (\omega/2) \sqrt{1-(2\gamma/\omega_d)}$. Following some algebraic manipulations, this result can be cast into
\begin{equation*}
c(s)=2c(0){\rm e}^{-\frac{w_{d}}{2}s}\left[\frac{w_{d}}{2\Omega}\sinh\left(\Omega s\right)+\cosh\left(\Omega s\right)\right],
\end{equation*}
from which we recover
\begin{equation*}
u(s,s')=\frac{c(s')}{c(s)}={\rm e}^{\frac{w_{d}}{2}(s-s')}\frac{\frac{w_{d}}{2\Omega}\sinh\left(\Omega s'\right)+\cosh\left(\Omega s'\right)}{\frac{w_{d}}{\Omega}\sinh\left(\Omega s\right)+\cosh\left(\Omega s\right)}. \nonumber
\end{equation*}

Since our goal is to construct contribution from this particular channel to the exponent of the mean entanglement $\bar{x}(t)$, function in Eq.~(10) of the main text, we need to calculate
\begin{equation}
\label{eq:Gamm}
\Gamma(s)=\int_{0}^{s}\alpha(s-s')u(s,s')ds',
\end{equation}
for this specific choice. Substitution of $\alpha(s)$ and $u(s,s')$ gives finally
\begin{equation}
\Gamma(s)=\frac{\frac{\omega_{d}}{2\Omega}\sinh\Omega s}{\frac{\omega_{d}}{2\Omega}\sinh\Omega s +\cosh \Omega s},
\end{equation}
thus completing the calculation.


\begin{thebibliography}{99}

\bibitem{zurek} Zurek,~W.~H. Decoherence, einselection, and the quantum origins of the classical. {\it Rev.~Mod.~Phys.} {\bf 75}, 715-775 (2003).

\bibitem{meas1} von~Neumann,~J. {\it Mathematical foundations of Quantum Mechanics} (Princeton, 1996). 

\bibitem{meas2} Kraus,~K. {\it States, effects, and Operations: Fundamental Notions of Quantum Theory} (Springer, Berlin, 1983). 

\bibitem{open1} Carmichael,~H.~J. {\it An Open Systems Approach to Quantum Optics} (Springer-Verlag, Berlin, 1993)

\bibitem{open2} Gardiner,~C.~W. {\it Quantum Noise} (Springer-verlag, Berlin, 1991).

\bibitem{nuev0} Caravalho,~A.~R.~R., Busee.~M., Brodier,~O., Viviescas,~C. and Buchleitner,~A. Optimal Dynamical Characterization of Entanglement. {\it Phys. Rev. Lett.}  {\bf 98}, 190501 (2007).  

\bibitem{nuev1} Viviescas,~C., Guevara,~I., Caravalho,~A.~R.~R., Busee,~M. and Buchleitner,~A. Entanglement Dynamics in Open Two-Qubit Systems Via Diffusive Quantum Trajectories. {\it Phys. Rev. Lett.}  {\bf 105}, 210502 (2010).  

\bibitem{nuev2} Vogelsberger,~S. and Spehner,~D. Average entanglement for Markovian quantum trajectories {\it Phys. Rev. A} {\bf 82}, 052327 (2010). 

\bibitem{qi} Mintert,~F., Caravalho,~A.~R.~R., K\'us,~M. and Buchleitner,~A. Measures and dynamics of entangled states. {\it Phys. Rep}, {\bf 415} 207 (2005). 

\bibitem{qi1} Buchleitner,~A., Viviescas,~C. and Tiersch,~M. (Eds.), {\it Entanglement and decoherence: Foundations and Modern Trends}, Lect. Notes in Phys. 768 (Springer, Berlin Heilderberg 2009).

\bibitem{scal2} Gour,~G. Evolution and Symmetry of Multipartite Entanglement. {\it Phys. Rev. Lett.} {\bf 105}, 190504 (2010).

\bibitem{scal4} Verstraete,~F., Dahaene,~J. and De Moor,~B. Normal forms and entanglement measures for multipartite quantum states. {\it Phys. Rev. A} {\bf 68}, 012103 (2003).

\bibitem{concu1} Wootters,~W.~K. Entanglement of Formation of an Arbitrary State of Two Qubits. {\it Phys. Rev. Lett.} {\bf 80}, 2245 (1998).

\bibitem{exp} Liu, B-H. {\it et.~al.} Experimental control of the transition from Markovian to non-Markovian dynamics of open quantum systems. {\it Nature Phys.} {\bf 7}, 931-934 (2011).

\bibitem{cond1} Werner,~R.~F. Quantum states with Einstein-Podolsky-Rosen correlations admitting a hidden-variable model. {\it Phys. Rev. A} {\bf 40}, 4277 (1989).

\bibitem{cond2} Uhlmann,~A. Fidelity and concurrence of conjugated states. {\it Phys. Rev. A} {\bf 62}, 032307 (2000).

\bibitem{scal1} Konrad,~T., de Melo,~F., Tiersch,~M., Kasztelan,~C., Aragao,~A. and A.~Buchleitner, A factorization law for entanglement decay. {\it Nature Phys.} {\bf 4}, 99 (2008), 

\bibitem{scal3} Jimenez-Farias,~O., Lombard-Latune,~C., Walborn,~S.~P., Davidovich,~L. and Souto-Ribeiro,~P.~H. Determining the dynamics of entanglement. {\it Science} {\bf 324}, 1414-1417 (2009).  

\bibitem{nuev2.5} Verstraete, F., Wolf, M. M. and Cirac, J. I. Quantum computation and quantum-state engineering driven by dissipation. {\it Nature Phys.} {\bf 5}, 633-636 (2009).

\bibitem{nuev3} Mascarenhas,~E., Cavalcanti,~D., Vedral,~V. and Franca-Santos,~M. Physically realizable entanglement by local continuous measurements. {\it Phys. Rev A.} {\bf 83}, 022311 (2011). 

\bibitem{nuev4} Caravalho,~A.~R.~R. and Franca-Santos,~M. Distant entanglement protected through artificially increased local temperature. {\it New. Jour. Phys.} {\bf 13}, 013010 (2011).

\bibitem{nuev5} Franca-Santos,~M., Terra Cuhna,~M., Chaves,~R. and Caravalho,~A.~R.~R. Quantum computing with incoherent resources and quantum jumps. {\it quant-ph} 111.1319 (2011).

\bibitem{nuev6} Zurek.~W.~H. Quantum Darwinism {\it Nature Phys.} {\bf 5}, 181-188 (2009).

\bibitem{qt0} Diosi,~L. and Strunz,~W.~T. The non-Markovian Schr\"odinger equation for open systems. {\it Phys. Lett. A}  {\bf 235}, 569-573 (1997).

\bibitem{qt1} Diosi,~L., Gisin, N. and Strunz,~W.~T. Non-Markovian Quantum State Diffusion. {\it Phys. Rev. A}  {\bf 58}, 1699-1712 (1998).

\bibitem{qt2} Strunz,~W.~T., Diosi,~L. and Gisin,~N. Open System Dynamics with Non-Markovian Quantum Trajectories. {\it Phys. Rev. Lett.}  {\bf 82}, 1801-1805 (1999).

\bibitem{rmt1} Popescu,~S., Short,~A.~J. and Winter,~A. Entanglement and the foundations of statistical mechanics. {\it Nature Phys.} {\bf 2}, 754-758 (2006).  

\bibitem{rmt2} Szarek,~S., Werner,~E. and {\.Z}yczkowski,~K. How often is a random quantum state k-entangled? {\it J. Phys. A} {\bf 44}, 045303 (2011).
 
\bibitem{ESD1} {\.Z}yczkowski,~K., Horodecki,~P., Horodecki,~M. and Horodecki,~R. Dynamics of quantum entanglement. {\it Phys. Rev. A} {\bf 65}, 012101 (2001).

\bibitem{ESD2} Simon,~C. and Kempe,~J. Robustness of multipartite entanglement. {\it Phys. Rev. A} {\bf 65}, 052327 (2002).

\bibitem{ESD3} D\"ur,~W. and Briegel,~H.~J. Stability of Macroscopic Entanglement under Decoherence. {\it Phys. Rev. Lett.} {\bf 92}, 180403 (2004).

\bibitem{ESD4} Yu,~T. and Eberly,~J.~H. Sudden death of entanglement. {\it Science} {\bf 323}, 555-558 (2009).

\bibitem{ESD5} Almeida,~M.~P., de Melo,~F., Hor-Meyll,~M., Salles,~A., Walborn,~S.~P., Souto Ribeiro,~P.~H. and Davidovich,~L. Environment-induced sudden death of entanglement. {\it Science} {\bf 316} 579-582 (2007).

\bibitem{ESD6} Yu,~T. and Eberly,~J.~H. Finite-Time Disentanglement Via Spontaneous Emission. {\it Phys. Rev. Lett.} {\bf 93}, 140404 (2004).

\bibitem{us} J.~D.~Urbina and C.~Viviescas, in preparation.

\bibitem{qt3} Strunz,~W.~T., Diosi,~L. and Gisin,~N. "Non-Markovian quantum state diffusion and open system dynamics" in Blanchard {\it et al} (Eds.) {\it Decoherence: Theoretical, Experimental, and Conceptual Problems}, Lect. Notes in Physics (Springer, Berlin Heildelberg 2000).

\bibitem{hus} Scully,~M.~O. and Zubairy,~M.~S. {\it Quantum Optics} (Cambridge University Press, 1997).

\end{thebibliography}
\end{document}